\documentclass[a4paper,12pt,useAMS,numberedappendix]{emulateapj}

\usepackage{graphics,graphicx,rotating}
\usepackage{natbib}
\usepackage{xspace}
\usepackage{amssymb}
\usepackage{amsmath}
\usepackage{subfigure}

\usepackage{color}
\usepackage{soul}
\usepackage{amsmath}
\usepackage{multirow}
\usepackage{booktabs}

\usepackage{natbib}

\newcommand{\simgt}{\lower.5ex\hbox{$\; \buildrel > \over \sim \;$}}
\newcommand{\simlt}{\lower.5ex\hbox{$\; \buildrel < \over \sim \;$}}
\def\balpha{\mbox{\boldmath $\alpha$}}
\def\bbeta{\mbox{\boldmath $\beta$}}
\def\btheta{\mbox{\boldmath $\theta$}} 
\def\bnabla{\mbox{\boldmath $\nabla$}}

\def\br{\mbox{\boldmath $r$}}
\def\bv{\mbox{\boldmath $v$}}
\newcommand{\FLASH}{{\sc FLASH}}

\newcommand{\rmsub}[1]{\ensuremath{_{\rm #1}}}
\newcommand{\RVIR}{{\ensuremath{R\rmsub{vir}}}}

\newcommand{\nop}{{\noindent}}

\newcommand{\KMSEC}{{$\rm km\,s^{-1}$}}
\newcommand{\MSEC}{{$\rm m\,s^{-1}$}}
\newcommand{\CMSEC}{{$\rm cm\,s^{-1}$}}

\newcommand{\HMPC}{{$\rm h^{-1\,}$Mpc}}

\newcommand{\MSUNFOUR}{{$10^{\,14}\,$\ensuremath{\mbox{\rm M}_{\odot}}}}
\newcommand{\MSUNFIVE}{{$10^{\,15}\,$\ensuremath{\mbox{\rm M}_{\odot}}}}
\newcommand{\TMSUNFOUR}{{$\times 10^{\,14}\,$\ensuremath{\mbox{\rm M}_{\odot}}}}

\newcommand{\simless} 
     {\ensuremath{\lower 3pt\hbox{$\rlap{\raise5pt\hbox{$\char'074$}}\mathchar"7218$}}}
\newcommand{\simgreat}
     {\ensuremath{\lower 3pt\hbox{$\rlap{\raise5pt\hbox{$\char'076$}}\mathchar"7218$}}}

\newcommand{\DEG}{{$^{\circ}$}}
\newcommand{\ASEC}{\ensuremath{\arcsec}}
\newcommand{\AMIN}{\ensuremath{\arcmin}}

\newcommand{\LCDM}{{\sc $\Lambda$CDM}}

\newcommand{\CHANDRA}{{\sc Chandra}}

\newcommand{\HST}{{\sc HST}}

\lefthead{Molnar et al.}
\righthead{Tangential Velocity of the Dark Matter in the Bullet Cluster}

\begin{document}

\title{
Tangential Velocity of the Dark Matter in the Bullet Cluster from Precise Lensed Image Redshifts
}

\author{
Sandor M. Molnar\altaffilmark{1}, Tom Broadhurst\altaffilmark{2,3}, Keiichi Umetsu\altaffilmark{4}, 
Adi Zitrin\altaffilmark{5}, Yoel Rephaeli\altaffilmark{6,7}, Meir Shimon\altaffilmark{6}
} 

\altaffiltext{1}{Leung Center for Cosmology and Particle Astrophysics, 
                      National Taiwan University, Taipei 10617, Taiwan; sandor@phys.ntu.edu.tw}

\altaffiltext{2}{Fisika Teorikoa, Zientzia eta Teknologia Fakultatea, Euskal Herriko Unibertsitatea UPV/EHU, 
644 Posta Kutxatila, 48080 Bilbao, Spain}

\altaffiltext{3}{IKERBASQUE, Basque Foundation for Science, Alameda Urquijo, 36-5, 48008 Bilbao, Spain}
               
\altaffiltext{4}
 {Institute of Astronomy and Astrophysics, Academia Sinica,
  P.~O. Box 23-141, Taipei 10617, Taiwan}
  
\altaffiltext{5}
  {Universit\"at Heidelberg, Zentrum f\"ur Astronomie, Institut f\"ur Theoretische Astrophysik, 
    Philosophenweg 12, 69120 Heidelberg, Germany}
                
\altaffiltext{6}
  {School of Physics and Astronomy, Tel Aviv University, Tel Aviv 69978, Israel}

 \altaffiltext{7}
 {Center for Astrophysics and Space Sciences, University of California, San Diego, La Jolla, CA, 92093, USA}

\begin{abstract}
    We show that the fast moving component of the ``bullet cluster'' (1E0657-56)
    can induce potentially resolvable redshift differences
  between multiply-lensed images of background galaxies. 
  This moving cluster effect, due to the tangential peculiar velocity of the lens,
  can be expressed as the scalar product of the lensing deflection
  angle with the tangential velocity of the mass components, and it is maximal for
  clusters colliding in the plane of the sky with velocities boosted by their mutual gravity. 
  The bullet cluster is likely to be the best candidate for the first measurement of this effect 
  due to the large collision velocity and because the lensing deflection and the cluster 
  fields can be calculated in advance.
  We derive the deflection field using multiply-lensed background galaxies detected with the 
  {\it Hubble Space Telescope}.
The velocity field is modeled using self-consistent {\it N}-body/hydrodynamical simulations
  constrained by the observed X-ray and gravitational lensing features of this system. 
  We predict that the triply-lensed images of systems ``G'' and ``H'' 
  straddling the critical curve of the bullet component will show the largest frequency shifts up
  to $\sim$0.5 \KMSEC.
  This is within the range of the {\it Atacama Large Millimeter/sub-millimeter Array} (ALMA)
  for molecular emission, and is near the resolution limit of the new generation high-throughput 
  optical-IR spectrographs.
  A detection of this effect measures the tangential motion of the subclusters directly, 
  thereby clarifying the tension with \LCDM, which is inferred from gas motion less directly.
This method may be extended to smaller redshift differences using the Ly-$\alpha$
 forest towards QSOs lensed by more typical clusters of galaxies.
More generally, the tangential component of the peculiar velocities of clusters derived 
by our method complements the radial component determined by the kinematic SZ effect,
providing a full 3-dimensional description of velocities.
\medskip
\end{abstract}


%
%
\section{Introduction}
\label{S:intro}

  The extreme physical conditions within galaxy clusters generate many
  observationally distinct phenomena, detected over the full spectral
  energy range. The ``bullet-cluster'' (1E0657-56), at redshift $z$ = 0.296,
  is one of the most energetic examples of clusters in collision displaying a large 
  supersonic gas Mach cone displaced by about 200 kpc from its parent dark
  lensing halo \citep{Clowe06,Bradac06}.
  This configuration, when modeled hydrodynamically, implies an impact velocity 
  of 3000 \KMSEC, seemingly exceeding the escape velocity of the system.
  This is quite unlike typical encounters expected in the hierarchical merging
  process, where typically only about 500--600 km s$^{-1}$ is predicted between
  massive clusters in the context of $\Lambda$ cold dark matter ($\Lambda$CDM; 
  \citealt{LeeKomatsu2010ApJ718,ThomNaga2012MNRAS419}), 
  so that clusters quickly merge after colliding.
  The probability of such a large velocity encounter in this context is
  very small, only about 3$\times 10^{-3}$ for merging clusters with 
  masses exceeding \MSUNFOUR\ is expected out to the redshift of the
  bullet cluster in the whole sky \citep{ThomNaga2012MNRAS419}, 
  with an upper limit of about 1900 \KMSEC\ predicted for the single most extreme
  encounter within this volume.

  Interestingly, the bullet cluster does not appear to be unique. 
  Several new examples of high speed gaseous bullets have been reported with
  gas velocities of $\simgt$ 2000 \KMSEC.
  X-ray and improved Sunyaev--Zel'dovich (SZ) effect 
  (\citealt{SZ1972CASP4}; for reviews, see \citealt{Birkinshaw1999PhR310,Carlstrom2002ARAA40})
  related measurements have uncovered examples of  
  large, about 100 kpc or greater, 
  displacements between the dark matter (DM)
  centers, X--ray and SZ peaks, suggesting massive high speed 
  cluster collisions (as shown by \citealt{Molnaret2012ApJ748}),
  some with large scale, in the order of 100 kpc, supersonic shocks 
   \citep{DawsonET2012ApJ747L,Russell2012MNRAS423,Menanteau2012ApJ748,
           Korngut2011ApJ734,AMIET2011MNRAS414,MertenET2011MNRAS417,RussellET2010MNRAS406}, 
  similar to that of the bullet cluster.
  This newly discovered phenomenon may 
pose a serious challenge to the concordance \LCDM, 
  and hence it is of great interest to pursue all possible independent
  estimates of the internal gas and DM motions,
  not only for these extreme examples, but also for peculiar motions of the
  cluster population in general, as a clear test of the viability of
  the \LCDM\ model.

 Any anomaly in this respect must be regarded as a very important clue
 to a more perfect physical understanding of cosmology in general, as
 the peculiar motions of clusters are not complicated by virialization
 and non--gravitational processes as in the case of galaxies which
 move within collapsed structures 
 (groups or clusters of galaxies) requiring 
 complicated modeling to interpret their relative motions 
 \citep{Peacocket2001Natur410,ReidET2012MNRAS426}.
 Instead, any anomalous cluster motion to emerge in a careful
 comparison with \LCDM\ would therefore imply the existence of some
 additional large scale forces for example \citep{FarrarRosen2007PhRvL98}, 
 or the consequence of self-interacting scalar fields 
\citep{HARKO11PhRvD83,MadTot12,Kain2012PhRvD85}.

  There are only few methods for direct determination of cluster peculiar velocities.
  The radial peculiar motion of clusters may be examined via accurate measurements
  of the doppler--shifted SZ effect, the kinematic SZ effect (kSZ; 
  \citealt{SZ1972CASP4}; for reviews, see \citealt{Birkinshaw1999PhR310,Carlstrom2002ARAA40}).
   This effect is due to inverse Compton scatterings between photons of the 
   cosmic microwave background (CMB) and the hot
   electrons in the intracluster gas having a bulk motion with respect to the universal CMB frame. 
  The kSZ effect may eventually be used to derive the distribution of cluster
  radial peculiar velocities, helpful in constraining cosmological models.

  Using statistical methods, \cite{HandET2012PhRvL109} found evidence for
  the kSZ effect in galaxy clusters based on data taken by the Atacama Cosmology Telescope (ACT).
  The first tentative detection of the kSZ effect was found for a massive component of
  the complex cluster MACS J0717.5+3745 using data from MUSTANG on the Green Bank Telescope (GBT)
  and Bolocam of the Caltech Submm Observatory (CSO) \citep{MroczkowskiET2012ApJ761}.
  We expect a statistical measurement of the kSZ effect may soon emerge from the
  Plank all--sky survey with the completion of several years of data
  \citep{Maket2011ApJ736}.
  However, the detection is challenging, and requires an accurate
  subtraction of the regular SZ signal, via multi-frequency SZ
  imaging, with ideally higher angular resolution, so that models for
  the gas distribution are less uncertain. 
  Internal bulk motions and
  turbulence will limit the precision of such estimates, though
  hopefully, detailed X-ray spectroscopy can explore this in
  principle by the use of the abundant X-ray emission lines. 
  The instrumentation to achieve this at a precision of about 100 \KMSEC\ 
  will hopefully be demonstrated with planned missions 
  (e.g., the International X-ray Observatory (IXO), see
  \citealt{Barconsetal201102.2845}).
  High resolution X-ray spectroscopy can also provide the depth of the
  gravitational potential via the gravitational redshift, which is
  estimated to be of order 50 \KMSEC\ in the core regions of massive
  clusters (\citealt{Broadhurst2000ApJ533}; 
  see also \citealt{WotjakET201109.6571}), 
  and must be taken into
  account when examining cluster motions based on gas emission.

  Direct detection of the motion of the cluster potential 
  free of the complication of gas hydrodynamics have been 
  also proposed and relate much more directly to the cosmological
  quantities of interest. The unscattered CMB is
  modified as it traverses the moving potential with a dipole
  pattern perturbation in frequency 
  due to the Birkinshaw--Gull effect 
  \citep{BirkGull83,GurvitsMitrof86,PyneBirkinshaw93}.
  This frequency shift is caused by the tangential
  component of the moving gravitational potential and originally derived as a
  lensing related phenomenon. It has subsequently been realized that
  this effect can be equivalently expressed as a simple geometric
  convolution of a suitably mass weighted line of sight tangential momentum of
  all the cluster material convolved with inverse angular separation,
  and may be regarded as a special case of the Rees-Sciama (RS) effect 
  \citep{Rubinoet2004AA419,Caiet2010MNRAS407}.

  In terms of the perturbation to the CMB, this tangential motion
  effect is several times weaker than the kSZ effect, and
  hence dominates only for clusters moving fast and close to the plane of the sky
  \citep{Rubinoet2004AA419,Caiet2010MNRAS407}.
Although this is not a scattered CMB signal, it has to be extracted from the same 
measurement database that includes also the thermal and kinematic SZ components.
  In principle, the differing angular signatures of these
  effects may be distinguished with further technical improvements
  allowing sufficiently detailed SZ maps to be constructed. The
  tangential signal is maximized for colliding clusters, moving close
  to the plane of the sky, accelerated by their mutual gravity. 
  A detection may be possible in the near future with the ongoing 
  South Pole Telescope\footnote{$http:/pole.uchicago.edu/$} (SPT) survey, 
  by coadding approximately $10^3$ merging clusters
  with averaging over the inherent temperature
  structure of the CMB and the complexities of intracluster gas
  motions \citep{Maturiet2007AA467}.

  Instead, the motion of the potential may be better examined in
  practice by measuring the difference in frequency shifts between
  multiple images of the same background source generated by 
  a tangentially moving cluster, as advocated by \citet{MB2003ApJ586}.
We will refer to this effect hereafter as the ``moving cluster effect'' to distinguish 
it form the BG effect describing the frequency shift in the CMB.
  Here the limitation comes from the intrinsic
  width of the spectral features used in the comparison between
  images, requiring bright sources and/or a distinctive internal
  velocity structure. Perhaps lensed QSOs offer the best hope for detecting
  this effect by utilizing the forest of narrow and numerous
  absorption lines. A few examples are known of relatively bright QSOs
  lensed by massive clusters, with the best cases
  being SDSSJ1004 \citep{SharonET2005ApJ629L,InadaET2003Natur426}, 
  SDSSJ2222+2745 \citep{DahleET2012arXiv1211.1091}, 
and SDSS J1029+2623 \citep{OtaET2012ApJ758,InadaET2006ApJ653L}, 
  with upcoming surveys such as the Hyper Suprime-Cam (HSC) 
  on the Subaru telescope expected to provide many more useful examples.

 In this context, the bullet cluster \citep{Markevitch2004ApJ606,Clowe06,Bradac06}
 is a compelling target because of its large tangential motion, combined with the
 large bending angles found for the multiply-lensed images identified
 around the two main mass components of this system 
 \citep{Mastropietro08,Bradac06}.
 The massive sub-cluster (the "bullet") with $M_{\rm{bullet}} \simeq 1.2 \times$\MSUNFOUR\
 is clearly on its way out of the cluster, separated presently by about 800 kpc 
 from the more massive main component, $M_{\rm{main}} \simeq 1.24 \times$\MSUNFIVE\
 \citep{Barrena2002AA386}, at a surprisingly high relative velocity \citep{Clowe04,Clowe06,Bradac06},
 which seems to exceed the escape velocity from the system by a wide margin.

 A clear measurement of the motion of the mass within the bullet
 cluster may help considerably in clarifying the anomalously large
 relative velocity of the pair of colliding clusters inferred for this
 system, for which X-ray based estimates vary considerably. The deep
 \CHANDRA\ data, where the bullet shaped gaseous object is visible,
 provide estimates of the shock Mach number and the pre-shock
 temperature to infer a shock velocity of $v_{\rm{shock}} = 4750^{+710}_{-550}$ \KMSEC\ 
\citep{Markevitch200511345,FarrarRosen2007PhRvL98}.

However, as it was pointed out by \cite{MIlos2007ApJL661} using a 2-dimensional (2D)
Eulerian code, \FLASH,
the relative velocity of the two main DM components can be substantially lower,
about 4000 \KMSEC, in the case of the bullet cluster. 3-dimensional
(3D) smoothed particle hydrodynamics (SPH) numerical simulations of \citet{Springel07}
showed that even though the shock velocity is 4750 \KMSEC, the relative velocities
of the two DM components are only about 3000 \KMSEC, because the
preshock intracluster gas of the main cluster is falling on the bullet by about 1100 \KMSEC,
and the shock is moving with a velocity of about 600 \KMSEC\ relative to the bullet DM.
\citet{Springel07} concluded that the initial velocity at a separation of 3.4 Mpc is only 
about 2000 \KMSEC. 
 On the other hand, \citet{Mastropietro08}, based on their 3D SPH simulations, 
 argued that the simulations of \citet{Springel07} do not successfully reproduce the
 relatively large observed displacement between X--ray Mach cone and the
 location of the corresponding DM component of the bullet, 
 for which they infer a higher infall velocity of $\sim$\,3000 \KMSEC, 
 at an initially larger separation of 5\,Mpc. 
 Here we also infer a relatively high relative velocity from
 our 3D adaptive mesh FLASH-based N--body/hydrodymanical simulations  
 \citep{Molnaret2012ApJ748} 
 designed to investigate the relationship between 
 gas and DM in colliding clusters, described in Section~\ref{S:FLASH}.

 Depending on the estimated relative infall velocity of the DM
 components in the bullet cluster (2000 or 4000 \KMSEC),
 there may be a considerable tension with the standard \LCDM\ model. 
 A first estimate of the probability of such cases, using the result of 
 \cite{Markevitch200511345}, was made by \citet{Hayashiwhite06} for the
 Millennium Run \citep{Springelet2005NATURE435} in a search for sub-clusters 
 moving with a relative velocity of 4500 \KMSEC. The limited volume of the simulation 
 (500 \HMPC)$^{3}$, contains only a few halos with masses comparable to
 the bullet cluster, but they conclude that about 1\% of clusters reach this level.
 Therefore, even this over estimated infall velocity seems to be in accord with 
 the concordance \LCDM\ model.
 In contrast, from the larger MICE simulations, \citep{Crocce2010MNRAS403},
 the analysis of \citet{LeeKomatsu2010ApJ718} implies that \LCDM\ 
 is excluded at the 99.9\% confidence level for the initial conditions \citet{Mastropietro08} 
 inferred for the bullet cluster. 
 The number of halos with large pairwise velocities is small so
 that this estimate relies on a Gaussian tail extrapolation, therefore it is quite uncertain.
 \cite{ThomNaga2012MNRAS419} carried out large scale cosmological simulations 
 with sufficient statistics
  to determine the high velocity tail of the probability distribution
 of the pairwise velocities. 
 They found that the probability of finding a cluster merger with impact velocity 
\simgreat\ 3000 \KMSEC\ in a concordance \LCDM\ model is $3 \times
10^{-8}$, and thus they concluded that the bullet cluster is incompatible with the
\LCDM\ model 
(assuming that the predictions of the binary merger simulations for the impact velocities 
are correct).

These latest results motivate us
to pursue a more direct measurement of the motion of the DM via the relativistic
shift in the frequencies induced by tangential motion.

 This paper is organized as follows.  In Section~\ref{S:BASIS} we
 briefly summarize the frequency shift generated by the tangential
 motion of a mass, its relation to the velocity field and to the
 lensing deflection angle field. In this section we also discuss the feasibility of 
 observing small relative velocities using high resolution spectroscopy.
We provide
 rough estimates of the S/N that may be expected for the velocity 
 shifts between the known multiple images 
 and we also discuss other potentially more sensitive measurements 
 for multiple lensed QSOs behind massive lenses.
 In Section \ref{S:FLASH} we model the velocity field of the bullet-cluster using
 a FLASH-based AMR code, which approximately reproduces the observed
 gas and DM morphologies.  
 In Section~\ref{S:LENSING} we describe the method we use to measure
 deflection fields from sets of multiple images and apply this to
 images taken by 
the {\it Hubble Space Telescope} (\HST) 
of both components of the bullet-cluster. 
 In section \ref{S:RESULTS} we generate frequency shift maps based on the 
 product of the velocity field and the lensing deflection field derived from the data.
 We also compare
 with the deflection field corresponding to the best fitting
 dynamical model, and estimate the expected frequency shifts between
 the components of all the known multiply-lensed images. 
 Section~\ref{S:RESULTS} contains our results and conclusions.
 Our final comments can be found in Section~\ref{S:FINAL}.
 Throughout this paper, we adopt a concordance \LCDM\ cosmology with
 $\Omega_{m}=0.3$, $\Omega_{\Lambda}=0.7$, and $h\equiv H_0/(100\,
 {\rm km\, s^{-1}\, Mpc^{-1}})=0.7$,  
 errors represent a confidence level of $68.3\%$ ($1\sigma$)
 unless otherwise stated.

%
%
\section{Tangential motion and Frequency Shift}
\label{S:BASIS}

 Several authors have examined the effects that may result from
 massive moving objects by very different arguments, starting with
 \cite{PyneBirkinshaw93} and leading now to a full understanding of
 the role of the moving potential and the relative translational
 motions of observer, source, and lens
 \citep{MB2003ApJ586,WucknitzSperhake2004PhRvD,Rubinoet2004AA419,Sereno2008PhRvD},
 the consequences for lensing and SZ--related measurements,
 as well as the possible limiting effect of
 intervening motions on the detection of cosmological drift
 \citep{KilledarLewis2010MNRAS402}.

 If we include the aberration from the motion of a lens then the usual
 thin--screen approximation for the gravitational deflection of light rays is modified so
 that the angular position $\btheta_I$ of a lensed image is related to 
 the angular position $\btheta_S$ of the intrinsic source for a mass
 moving purely in the radial direction with velocity $v_r$ away from the
 observer:
\begin{equation}
\label{E:lenseq_rad}
                                      \btheta_S = \btheta_I-(1-\beta_r) \bnabla\psi(\btheta_I)
,
\end{equation}
where $\beta_r=v_r/c$,
$\psi(\btheta_I)$ is the effective lensing potential, and 
$\balpha\equiv \bnabla\psi(\btheta_I)$ is the deflection field.
The effective lensing potential is
defined by the 2D Poisson equation as $\triangle
\psi(\btheta_I)=2\kappa(\btheta_I)$.
Here, the source term represents the lensing convergene,
$\kappa(\btheta_I)=\Sigma(\btheta_I)/\Sigma_{\rm crit}$,
with $\Sigma_{\rm crit}$ the critical surface mass density for
gravitational lensing.
As pointed out by \cite{MB2003ApJ586},
the radial motion of a lens can modify the estimated mass from lensing but
only by as little as $\sim 0.3\%$ for radial motions below
1000\,\KMSEC. 
A very precise alternative means of measuring cluster mass determined
independently of lensing would be needed to obtain radial motions this way.

In the case of pure tangential motion, $\bbeta_T={\bv}_T/c$, the effect
on the deflection angles comes in the next order, set by the special relativistic 
change in momentum in the tangential direction, with Lorentz factor $\gamma=\sqrt{1-\beta_T^2}$
\begin{equation}
\label{E:lenseq_tan}
           \btheta_S=\btheta_I-\gamma\balpha(\btheta_I)
,
\end{equation}
so that for sub--relativistic speeds
this effect is too small to be detected using measurements of
multiple image positions. 
As it is clear from this equation, the effect of
the tangential motion, $\bbeta_T$, on the angular position of lensed
images is second order in $\beta_T$.
 Also in second order is the
effect of radial motion, $\beta_r$, of the lens which we also can
ignore since the tangential motion of the lens induces a frequency shift which 
{\it is first order in velocity}. 
To a very good approximation
the frequency shift between lensed images of the same source, 
$\Delta_\nu(\btheta_I) \equiv \Delta{\nu(\btheta_I)} / \nu_0$, can be expressed as
\begin{equation}
\label{E:DELNU}
      \Delta_\nu(\btheta_I) = \bbeta_T(\btheta_I) \cdot \balpha(\btheta_I)
                                                  = \beta_T(\btheta_I)\, \alpha(\btheta_I)\cos{\phi}
,
\end{equation}
where $\phi$ is the angle between the the projected direction of motion of the lens, 
$\bbeta_T$, at the observed angular position $\btheta_I$, and the the direction of
the deflection field $\balpha(\btheta_I)$ at this location (Equation 9 of \citealt{BirkGull83}).

\cite{Rubinoet2004AA419} showed that this effect can be equivalently expressed 
 in terms of the 3D velocity field, integrating the mass weighted velocity along the
 line of sight (and summed over all components of the mass
 distribution, DM and gas), so the frequency shift can be calculated as
\begin{equation} \label{E:integral}
        \frac{\Delta\nu(\br_T)}{\nu_0} = -\frac{4G}{c^3}\,\int\!d^3r'\,
                                                            \frac{ \rho(\br')\bv(\br')\cdot (\br_T-\br'_T)}{|\br_T-\br'_T|^2}
.
\end{equation}
From this, one easily sees that the RS effect measures the convergence
of the line off sight (LOS) momentum.

 Here we have considered only the tangential component of the motion
 of the lens, but in addition to this frequency shift, there will be
 frequency shifts from the relative peculiar motions of the observer and
 the source, which have been discussed in the context of CMB related
 effects by \cite{WucknitzSperhake2004PhRvD} and \cite{Sereno2008PhRvD}.
 Only simple doppler shifts are generated by gravitational lensing from 
 the radial components of the velocity of the observer and the source, and
 since we measure only relative frequency differences, therefore 
 these frequency shifts cancel out because they are the same for all images.
 Frequency shifts due to the tangential motion of the observer and the source 
 were studied in detail by \cite{WucknitzSperhake2004PhRvD}.
 They found that these contributions are linear in the tangential velocities, and
 can be expressed as
\begin{equation}    \label{E:WSFRSHIFT}
       \Delta_\nu (\btheta_I) = \left[  \bbeta_T^{\,L} - \frac{D_{LS}}{D_{OS}} \bbeta_T^{\,O} 
                      - \frac{D_{OL}}{D_{OS}} \frac{1 + z_L}{1 + z_S} \bbeta_T^{\,S} \right]   \cdot \balpha
,
\end{equation} 
where all vectors on the right hand side are evaluated at the image position, $\btheta_I$,
$D$ denotes angular diameter distance, and the indices, 
$O$, $L$, and $S$ refer to the observer, lens, and source
(Equation 37 of \citealt{WucknitzSperhake2004PhRvD}).
Note that the contribution to the frequency shift from the tangential motion of the
observer and the source are weighted down relative to that of the lens by the ratios
of the respective distances, and in the case of the source, also 
with an additional factor of the redshift ratio.
Therefore, we expect that, since the peculiar velocities of the
field galaxies, especially at large redshifts, are small, and the peculiar velocity of
the observer (our Heliocentric velocity relative to the CMB) is only 
$c \beta^{\,O} = v^O$ = 369.0$\pm$0.99 \KMSEC\ 
\citep{HinshawET2009ApJS180}, 
we can neglect this complication for the massive interacting clusters of interest here,
and use Equation~\ref{E:DELNU} for our calculations.

The maximum frequency change due to the moving cluster effect generated by
a cluster with a well--defined Einstein radius, $\theta_{\rm Ein}$, occurs
with a pair of multiple images $(i,j)$ displaced in the direction of motion of the 
cluster with respect to the universal CMB frame, when the difference
in the cosine term is at its maximum, $| \cos{\phi}_i - \cos{\phi}_j | = 2$, and 
\balpha\ = $\theta_{\rm Ein}$ in Equation~\ref{E:DELNU}. 
In this case, the difference in the frequency shift can be approximated as
\begin{equation}
\label{E:DNUEINST}
           \Delta_\nu(|\btheta_I^i-\btheta_I^j|) = \frac{ \left|\Delta\nu_{ij}\right|} {\nu_0}
                                                                      \simeq 2{v_T\over{c}}{\theta_{\rm Ein}}
.
\end{equation}
In terms of redshift difference for a source at $z_s$:
\begin{equation}
        |\Delta z_{ij}| = 3\times 10^{-6} (1+z_s) \left( \frac{v_t}{3000\,{\rm km\,s^{-1}}} \right)
                                                                                 \left(\frac{\theta_{\rm Ein}}{30\arcsec}\right),
\end{equation}
which corresponds to a velocity difference of 
$\Delta V\sim 1.5\,$km\,s$^{-1}/(1+z_s)$,
or a wavelength separation of $\Delta \lambda \sim 0.03\AA/(1+z_s)$ at $\lambda=8800\AA$.
In practice, the Einstein radius $\theta_{\rm Ein}$ of a massive cluster
can be constrained to within 
$\sigma_{\rm Ein}/\theta_{\rm Ein}\sim 10\%$ 
from detailed strong lens modeling
\citep[e.g.,][]{,Zitrinet2011MNRAS410,UmetsuET2012ApJ755}.

Assuming a signal-to-noise ratio per line of the order unity 
$\Delta_\nu/\sigma(\Delta_\nu)\sim 1$ and
combining $N$ such independent spectral lines, 
the net sensitivity for cluster peculiar velocity measurements can be
improved as
\begin{equation}
    \sigma(\beta)/\beta =\sqrt{\sigma_{\rm Ein}^2/\theta_{\rm Ein}^2 + \sigma^2(\Delta_\nu)/\Delta_\nu^2/N}
.
\end{equation}
Contamination in the deflection angle due to strong lensing
can be included in estimating $\sigma_{\rm Ein}/\theta_{\rm Ein}$.
The two main sources of contamination in the deflection angle                                                                             
are internal structure of the lensing subcluster and large-scale structure                
along the LOS.                                                                        
Our strong lensing method identifies and models the detectable lensing 
signals of the more massive cluster galaxies. 
Since the mass of each of these galaxies is less than $\sim 1\%$ of the                   
subcluster mass, roughly $\sim 1\%$ contamination would be expected from     
each of these galaxies.           
However, this uncertainly is already included in the 10\% error 
we quoted in the previous paragraph as deflection angle error.
There are more numerous smaller mass concentrations (galaxies)
in the subcluster which we can not identify. 
We can safely ignore their contribution to the error budget of the
deflection angle because their typical mass is less than $\sim 0.1\%$ of the 
subcluster mass, and due to their random distribution along the LOS to the lensed 
background galaxy, these are expected to contribute less than $\sim 1\%$ overall 
uncertainly in the deflection angle.       
Assuming that the tangential velocity is of order the velocity dispersion
of massive clusters, 1000 \KMSEC, a 1\% change in the deflection angle causes only a 0.25\% 
variation in the frequency change (Equation~\ref{E:DELNU}), 
and therefore this contamination can be ignored.

The contamination in the deflection angle from large-scale mass concentrations            
along the LOS is more important.
On large angular scales, $\sim$~2\DEG, the deflection due to lensing 
by the large scale structure is estimated to be $\sim$~2.5\AMIN\
(e.g., \citealt{PLANCK2013XVII}). However this deflection is
coherent, thus all the images of the subcluster and lensed 
background galaxies will shift in direction by the same amount; therefore, 
this contamination can be ignored.            
On arc second scale the density fluctuations along the LOS cause a change
in the deflection angles for background galaxies located at a redshift of $\sim$~2 
(the multiple lensed galaxies for the bullet range from $z$ = 1.3--2.1),
of $\sim$~5\% (smaller for $z \le 2$), but these shifts may also be correlated
(see Figure 1 of \citealt{Host2012MNRAS420}). 
This contamination has to be included in the uncertainly of the deflection angle. 
Adding in quadrature this 5\% uncertainty to the 10\% due to effects discussed 
above we obtain an uncertainly of 11\%. 
Assuming $N \approx 10$ for bright lensed QSOs, $\sigma(\Delta_\nu)/\Delta_\nu \approx 1$, 
and $\sigma_{\rm Ein}/\theta_{\rm Ein} \approx 0.11$, 
${\rm S/N}\equiv \beta/\sigma(\beta) \approx 3$.
An 11\% change in the deflection angle, assuming that the tangential velocity 
is of order the observed dispersion in the galaxy peculiar velocity, 600 \KMSEC\ 
\citep{RaySaslaw1996ApJ564}, 
results in only 1.65\% deviation in the frequency change.
However, this velocity dispersion includes member galaxies, which have large
peculiar velocities, therefore it is an overestimate of this effect.
We may consider this 1.65\% contamination in the frequency change as an upper limit, 
and therefore it can be ignored.

Currently the best velocity precision for faint galaxies has been
established empirically to be approximately 1 \KMSEC\
in terms of the
centroid of emission lines of lensed star forming galaxies with
typical velocity dispersions of 50 \KMSEC\
using the X-shooter spectrograph on the VLT 
\citep{Christensenet2010MNRAS406}.

In the case of lensed QSOs the numerous available Lyman forest lines
have line widths less than half that of faint galaxies so 
the error can be further reduced
by the square root of the number of absorption
lines, which run into many hundreds for well resolved
spectroscopy. With the new ALMA\footnote{$http://www.almaobservatory.org/$}.
array the required resolution can be readily achieved
at sub--mm frequencies. For ALMA it is important to
establish the lensed source redshifts accurately (through optical-IR
spectroscopy) in advance so that the 
respective frequency window corresponding
to molecular emission can be set in advance. This will
require optical-IR spectroscopy of only modest resolution (10\AA),
typical of many high throughput spectrographs.

There are only a few
good examples of QSOs lensed by clusters known to date, all discovered in
the SDSS survey, and many more examples will be found of fainter QSOs
lensed by clusters with the new Subaru/HSC survey and others.
Therefore, for the most promising colliding clusters we 
will have to rely on lensed galaxy images.
 In this case it is important to observe multiple images
for which there is little differential magnification across the galaxy
so the velocity profile is not skewed by this, making the smaller
bright multiply-lensed images lying away from the critical curves the
best targets.

We may hope to see resolved velocity features within the overall line profile, 
which will give us an opportunity to measure the cluster motion with 
greater precision
by identifying repeated internal velocity structures between
multiple-images of each source.
In general, we can improve this pairwise measurement for lensed sources 
with $N_i$ multiple images (where $i$, is the source index, $1 \le i \le P$, 
assuming $P$ lensed background galaxies) 
amounting to
$\Sigma_1^P ( N_i -1)$ independent relative frequency shifts measurements.

In the case of relaxed clusters, we can 
assume that the cluster peculiar velocity field $\bbeta$ is
uniform across
the Einstein radius (where multiply-lensed images are distributed), 
i.e., the cluster simply drifts through the rest frame of the CMB with a velocity 
independent of position. In this case
the pairwise frequency shifts of a given multiply-imaged source, 
$\Delta_{\nu}^{ij}  \equiv  \Delta_\nu(\btheta_I^i)-\Delta_\nu(\btheta_I^j)$,
at a source position $\btheta_S$ 
may be expressed (to first order in velocity) as 
\begin{equation}  \label{E:SHIFTS3}
       \Delta_{\nu}^{ij} = \bbeta_T\cdot (\balpha(\btheta_I^i)-\balpha(\btheta_I^j)) =
                                     \bbeta_T\cdot (\btheta_I^i-\btheta_I^j)
, 
\end{equation} 
where we have used Equation (\ref{E:lenseq_tan}) in the 
second equality and neglected higher-order terms.  
Since the image positions $\btheta_I$s of a lensed source are
direct observables, one can reconstruct the tangential cluster
velocity $\bbeta_T$ ($\beta_T$ and $\phi$) in a model--independent
manner for a lens system with $N\ge 3$ having precise frequency shift measurements, 
since no deflection field model is required. 
We expect that the peculiar velocities of relaxed clusters are smaller than
the components in merging clusters, but the statistic can be improved
if we have more sources with more than two multiple images. 
This work can be extended to a sample of relaxed clusters
to obtain their statistical bulk motions for comparison
with the theoretical expectations, and complementing such efforts
being made via current kSZ measurements.

 The cosmological expansion has a smaller effect on the redshift
 difference between a pair of images ($i,j$).
 It is given by the time delay $(\Delta t)_{ij}$ between the images, 
 producing effectively a small change in redshift due to the expansion of the
 universe, given by
\begin{equation}
\label{E:dzexpand}                           
                       (\Delta z)_{ij}=(1+z_{\rm lens}) H(z_{\rm lens})(\Delta t)_{ij}
.
\end{equation}
 Since empirically time delays of about 2--3 yrs are large between lensed
 cluster images and measured for QSOs 
 (\citealt{FohlmeisterET2013ApJ764} for SDSS J1029+2623 and 
  \citealt{FohlmeisterET2007ApJ662} for SDSS J1004+4112), 
 depending on the geometry, so that the typical redshift difference
 caused by the universal expansion during this time interval is,
 $\Delta z/(1+z)\sim 2\times 10^{-10}$ in 3\,years.
 This corresponds to a
 redshift difference of $cz/(1+z)\sim 6$ \CMSEC\ and so about 4 orders of magnitudes
 smaller than the redshift difference caused by the tangential motion,
 and hence can be completely ignored for our purposes.

 The above treatment constitutes a reasonable semi-quantitative description of the 
 several effects due to  
 a moving lens, but to calculate in a realistic case where
 we have an asymmetric deflection field derived from lensing or a
 complicated velocity field from a numerical simulation as in the case
 of the bullet cluster, 
 we should examine more
 general expressions to gain a better understanding of the 
 precision required to measure the frequency shifts for the actual
 multiply-lensed images detected in clusters of interest.

%
%
\begin{figure}
\centerline{
\includegraphics[width=.50\textwidth]{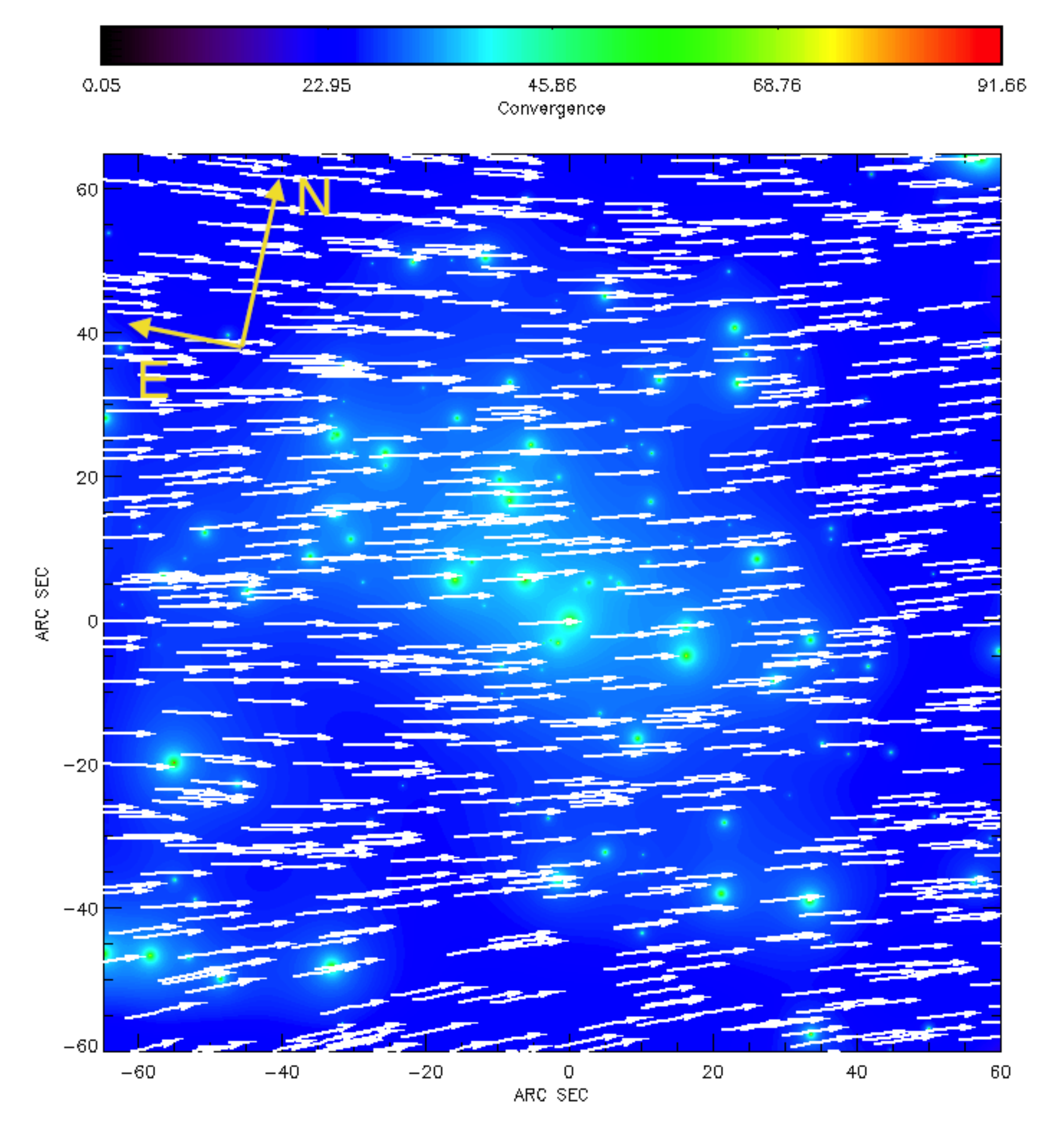}
}
\caption{
\protect
 False color image of the lensing convergence, $\kappa$,
 derived from ACS/\HST\ strong-lensing observations
 centered on the bullet in the bullet cluster with the 
 velocity field of the dark matter component of the bullet from our 
 best fit \FLASH\ simulation (shown in Figure~\ref{F:SIM_KAPVEL})
 overlaid. The color bar on top represents the scale of the convergence.
We used $S$ = 8 for the degree of the polynomial for smoothing
derived from our model fitting.
 The image coordinates are given in arcsec as derived from
 observations.
 The center of the coordinate system is the BCG of the bullet.
\medskip
\label{F:KAPVEL}
}
\end{figure} 

%
%
\section{Deflection field of the Bullet Cluster from strong lensing}
\label{S:LENSING}

In our analysis of the bullet cluster, we adopt our established method to
strong-lens modeling using multiple images of background galaxies visible in the
\HST/Advanced Camera for Surveys (ACS) data as described in
\cite{ZitrinET2009MNRAS396}.
Our previous work has uncovered a large number of
multiply-lensed galaxies in \HST\ images of many clusters,
including highly irregular merging systems for which the identification of
multiple images had remained unsolved due to the complexity of the
strong lensing regime requiring initially well-guessed models to guide the 
secure identification of counter images (e.g., \citealt{Zitrinet2011MNRAS410}).
The basic assumption of our method is that mass approximately traces light,
so that the photometry of the red cluster member galaxies is used as the
starting point for the model. 
We identify member galaxies using \HST\ multiband photometry
based on their position in color space close to the cluster sequence.
We model the mass surface density distribution of individual cluster galaxies
at angular position $\btheta$
assuming a symmetric power-law mass profile:
$\Sigma (\btheta) \propto (\btheta - \btheta_i)^{-q}$, where $\btheta_i$ is the
angular center of the $i$th galaxy, and $q$ is the power law index with the
amplitude scaled linearly with flux.
The deflection angle due to the galaxy component is the sum of all
individual galaxies, which in our case can be expressed as
\begin{equation}
    \balpha_{\rm gal}(\btheta) = A(q) \sum_i F_i \, | \btheta - \btheta_i |^{1-q} \,
                                                                      \frac{\btheta - \btheta_i}{| \btheta - \btheta_i |}
,
\end{equation}
where the sum is over all cluster galaxies, $F_i$ is the flux, and the                    
amplitude $A$ depends on $q$ and physical constants.                       
                                                             

%
%
\begin{figure}
\centerline{
\includegraphics[width=.50\textwidth]{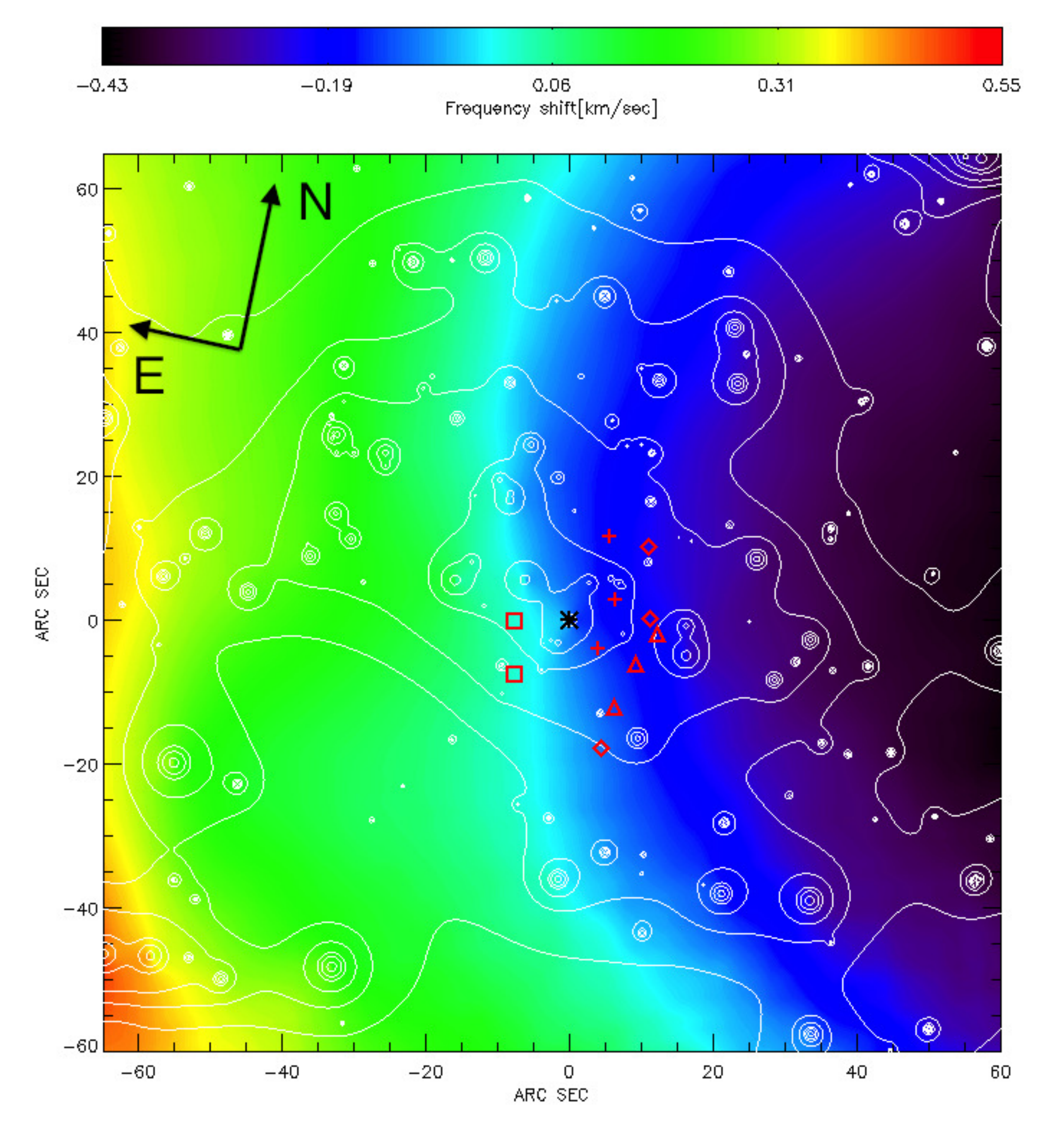}
}
\caption{
\protect
 Same as Figure~\ref{F:KAPVEL}, but for the frequency shift due to gravitational lensing 
 (reconstructed from strong lensing observations) and tangential motion of the bullet 
 (from our best fit simulation shown in Figure~\ref{F:SIM_KAPVEL}).
 The color bar on top represents the scale of the frequency shift in velocity units (\KMSEC).
 The white contours represent the convergence field shown in Figure~\ref{F:KAPVEL}
 in colors. 
The plus signs, diamonds, triangles, and squares represent the positions of
multiple lensed background galaxies (systems G, H, I, J) listed in Table~\ref{T:TABLE1}.
The star at the center marks the BCG of the bullet component.
\label{F:DVELKMS}
}
\end{figure} 

The mass surface density distribution of the cluster DM component is approximated 
by smoothing the resulting mass surface density of the cluster galaxy component
using a polynomial of degree $S$, which is taken as a free parameter.
We derive the deflection angle of this smooth component, $\balpha_{\rm DM}$,
representing the mass distribution at each pixel with a delta function located
at the pixel center:
\begin{equation}
\label{deflection_xDM}
            \alpha_{DM} (\btheta) = A(q) \sum_{i} m_i\,  \frac{ \btheta - \btheta_i }{ ( \btheta - \btheta_i )^2}
,
\end{equation}
where $m_{i}$, the mass of the $i$th cluster galaxy, represents the (unnormalized) 
mass value in the $i$th pixel.
Since the smooth component does not perfectly follow the light,
we assume an additional component for our lensing model to allow for further flexibility
in the form of a coherent spin-2 external shear, $\left(\Gamma^{\rm x}\right)_{k l}$ ($k$, $l$ = 1, 2).
The $k$th component of the deflection angle for this term therefore becomes:
$\alpha_{{\rm x},k}(\btheta) = \left(\Gamma^{\rm x}\right)_{k l} \Delta \theta_l$,
where $\Delta\btheta$ is the displacement vector of the
angular position $\btheta$ with respect to a fiducial reference position.
The amplitude, $|\gamma_{\,\rm x}|$, and the direction, $\phi_{\,\rm x}$,
of the external shear are also free parameters in our mass model.

Our model for the deflection field, $\balpha(\btheta)$, is thus a sum of
three components: the galaxy, $\balpha_{\rm gal}(\btheta)$, the smooth DM,
$\balpha_{\rm DM}(\btheta)$, and a component
due to the external shear, $\balpha_{\rm x}(\btheta)$,
weighted as $K$, 1 - $K$, and 1 respectively:
\begin{equation}
     \balpha(\btheta) = K \balpha_{\rm gal}(\btheta) + (1-K) \balpha_{\rm DM}(\btheta)
                                    + \balpha_{\rm x}(\btheta)
,
\end{equation}
where the weight, $K$, is our 5th free parameter.
Including an overall normalization factor, ${\cal N}$, this model
has 6 free parameters: ${\cal N},K,q,S,|\gamma_{\rm x}|,\phi_{\rm x}$.
As we have demonstrated in a series of papers
\citep{BroadhurstET2005ApJ621,ZitrinET2009MNRAS396,Zitrinet2011MNRAS410,
Zitrinet2011MNRAS413,ZitrinET201103.5618},
this approach to strong lensing is sufficient to accurately
predict the locations and internal structure of multiple images of background galaxies.

%
%
\begin{figure}
\centerline{
\includegraphics[width=.48\textwidth]{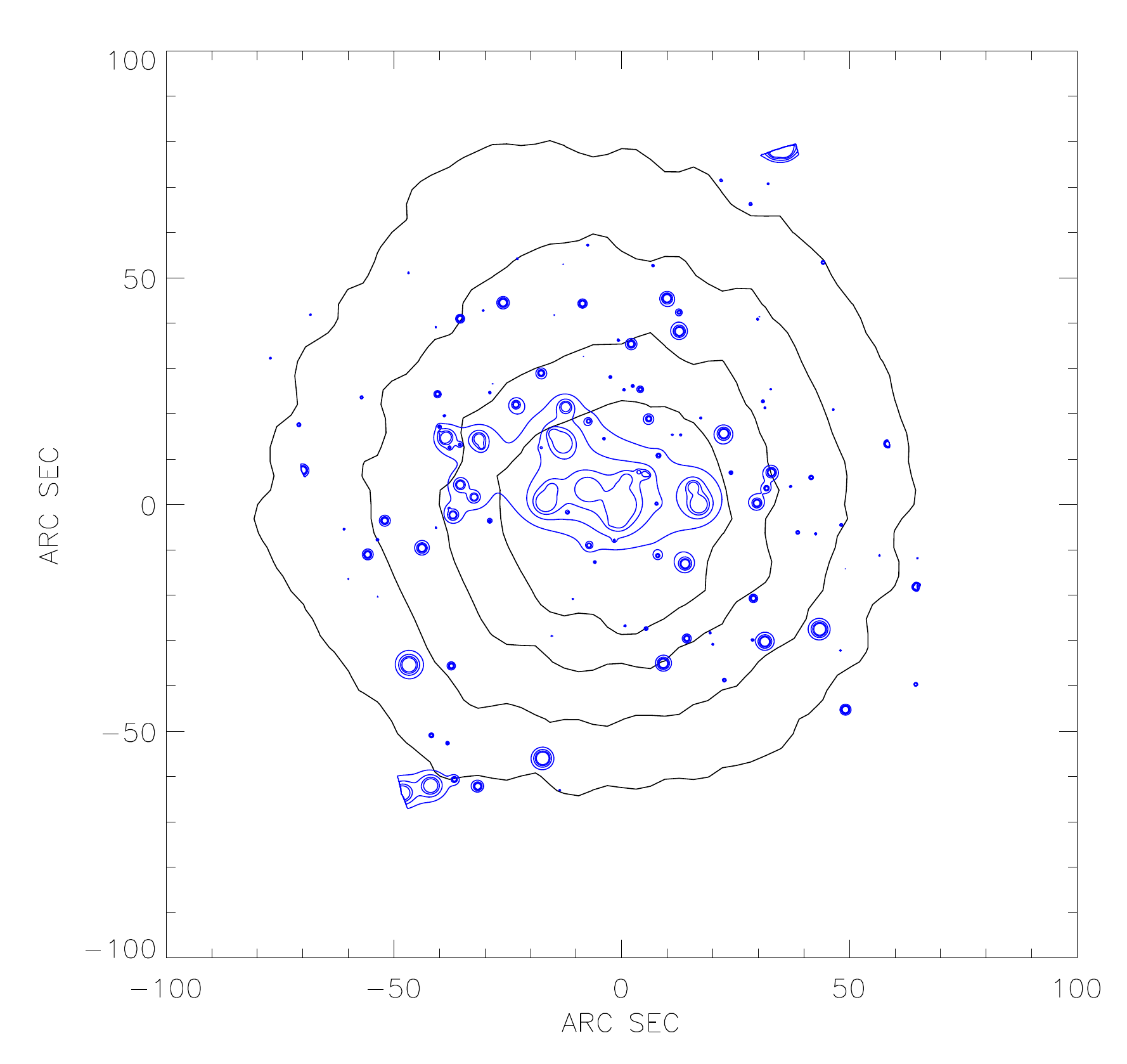}
}
\caption{
\protect
 Contours of convergence ($\kappa$ = 0.7,0.8,0.9,1.0)
 from strong lensing observations and our best fit \FLASH\ simulation, 
 shown in Figure~\ref{F:SIM_KAPVEL}, (blue and black contours) 
 centered on the BCG of the bullet (North is up, East is left).
\medskip
\label{F:KAPPACONT}
}
\end{figure} 

We derive our best-fit model using $\chi^2$ statistic, in particular,
we minimize the $\chi^2$ derived for the image plane:
\begin{equation}
\label{E:CHI2}
    \chi^2 = \displaystyle\sum_{i=1}^n {\frac{[\btheta_i -
                       \hat{\btheta}_i ({\cal N},K,q,S,|\gamma_{\rm x}|,\phi_{\rm x}) ]^2}{\sigma_i^2}}
,
\end{equation}
where $i$ runs from 1 to the number of lensed images, $n$,
$\btheta_i$ is the observed image position,
$\hat{\btheta}_i ({\cal N},K,q,S,|\gamma_{\rm x}|,\phi_{\rm x})$
is the image position given by our model,
and $\sigma_i$ is the measurement error in the position of the $i$th lensed image.
For each model parameter, we estimate the $1\,\sigma$ uncertainty by
$\Delta\chi^2 =1$ in the 6-dimensional parameter space.
The uncertainties in the $\Sigma(\btheta)$
field are estimated by propagating the
errors of all our strong-lens model parameters.

We show in Figure~\ref{F:KAPVEL} the convergence field  of the bullet cluster
derived from our strong-lensing analysis of the \HST/ACS images.
The smoothing was performed with a resolution of 2\ASEC, and the degree of the polynomial 
used, derived from our fitting using Equation~\ref{E:CHI2}, was 8.
The color bar indicates the values of the convergence, $\kappa$.
The large peaks of the mass density field mark the positions of individual galaxies.
These peaks can be seen clearly in Figure~\ref{F:DVELKMS} as white contours
overlaid on the predicted frequency shift due to the moving cluster effect
(see Section~\ref{S:RESULTS}).

%
%
\begin{figure}
\centerline{
\includegraphics[width=.49\textwidth]{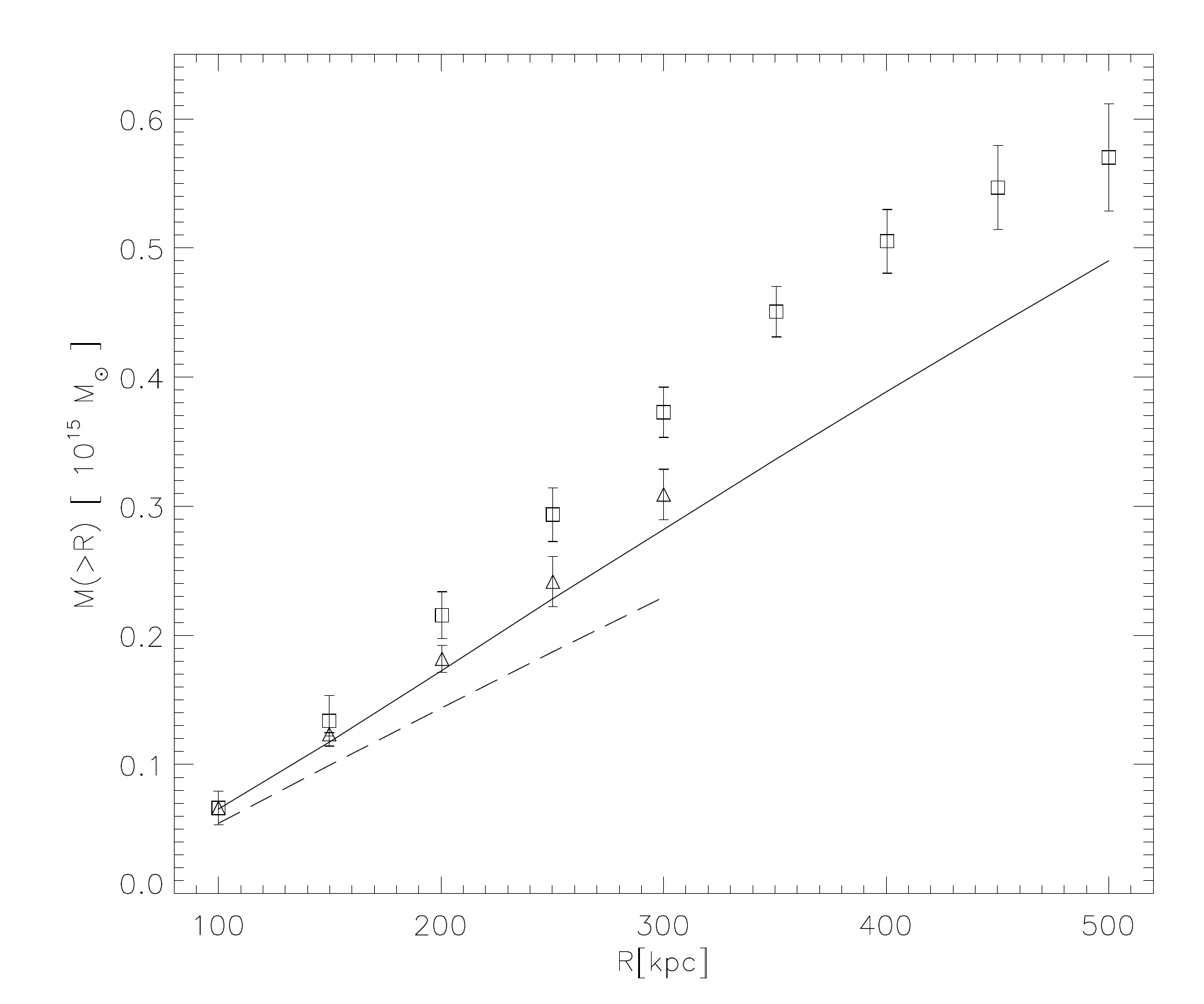}
}
\caption{
\protect
 Cumulative mass distribution for the two components of the bullet cluster.
 Squares and triangles with error bars represent masses of the main cluster 
 and the bullet derived from lensing observations \citep{Bradac06}, 
 solid and dashed lines show mass profiles from our best fit \FLASH\ simulation 
 (Figure~\ref{F:SIM_KAPVEL}) of the main cluster and the bullet.
\medskip
\label{F:MASSCUMUL}
}
\end{figure} 

%
%
\section{Velocity field of the Bullet Cluster from FLASH Simulations}
\label{S:FLASH}

Here we model the velocity structure of the
interacting bullet cluster, using the observational features visible
in this system, including the masses of the two cluster components and
their separation derived from lensing, and the gross
properties of the gas
determined from X-ray measurements.
We make use of the Eulerian adaptive mesh refinement parallel code, 
\FLASH\ developed at the Center for
Astrophysical Thermonuclear Flashes at the University of Chicago
\citep{Fryxell2000ApJS131p273}. 
We included the hydrodynamics module based on the Piecewise--Parabolic
Method (PPM) of \cite{1984JCoPh54p174}, and the 
$N$-body module with a multigrid solver \citep{2008ApJS176p293}
to represent DM particles.
We adopted a box size of 13.3 Mpc on a side 
so that we can follow all the matter for the duration
of the simulation.  
We reach a maximum refinement of 12.7 kpc at the high density 
(the cores of the mass components) and shocked regions.

%
%
\begin{figure}[t]
\centerline{
\includegraphics[width=.50\textwidth]{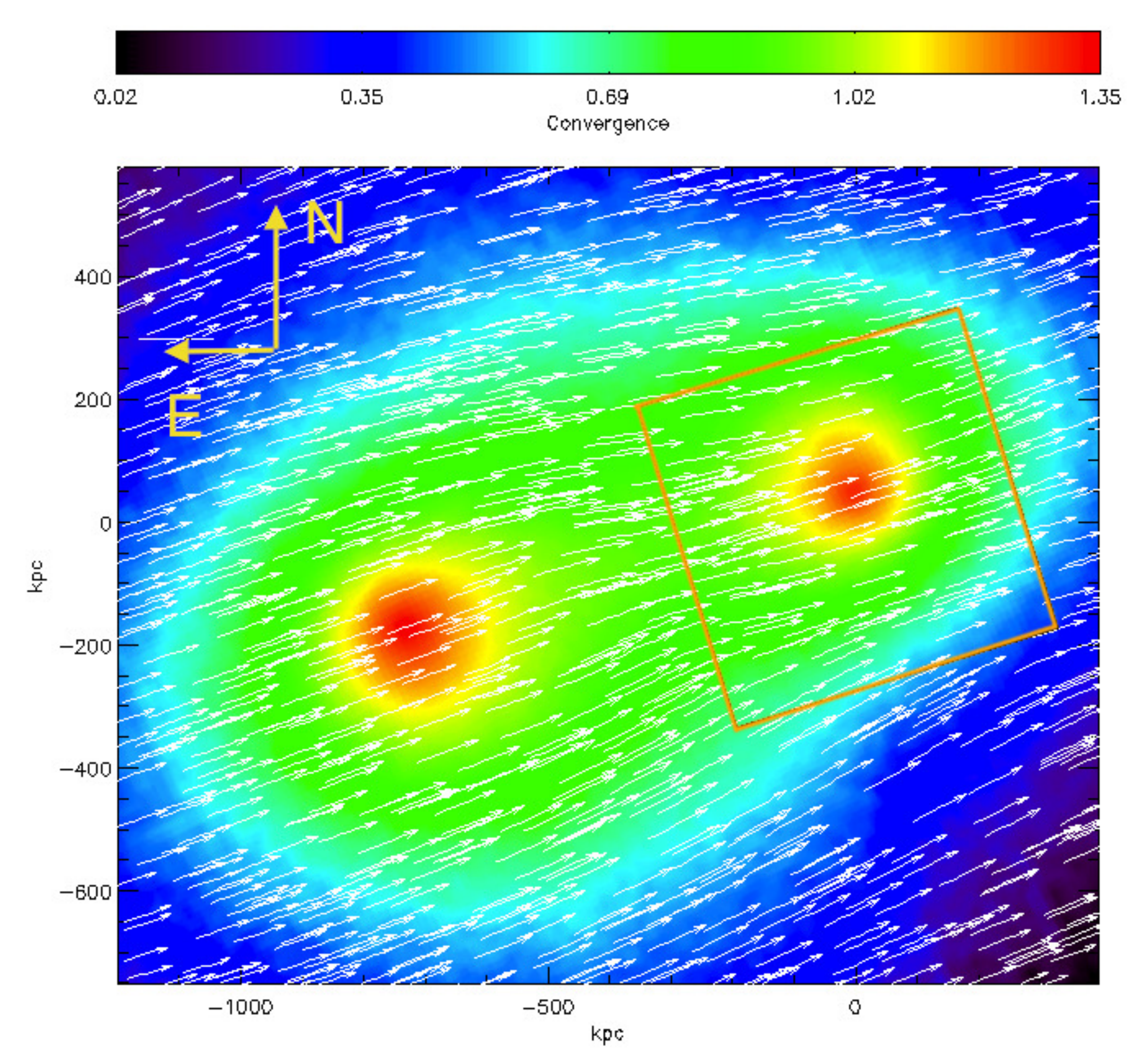}
}
\caption{
\protect
 False color image of the convergence with the velocity field 
of the bullet from our best fit \FLASH\ simulation overlaid (white arrows).
 The alignment of the image is the same as that of the bullet cluster
 (North is up, East is left).
The infalling cluster (the bullet; upper right) passed the main cluster 
(lower left) from the North moving toward North--West.
 The length of the white arrows represent the magnitude of velocity field
 with a maximum of about 4000 \KMSEC.  
 The orange box represents the area of the 
 ACS/\HST\ strong-lensing observations shown in Figure~\ref{F:KAPVEL}.
\label{F:SIM_KAPVEL}
}
\end{figure} 

We initially assume spherical cluster models with a cut off of the
distribution of the DM and gas at the virial radius, \RVIR\
\citep{BrNo98APJ495}.  
Within the virial radius, we assume an 
NFW model \citep{NFW1997ApJ490p493} for the DM density, 

\begin{equation}  \label{E:RHODM}
      \rho_{DM} (r) =  { \rho_s  \over x (1 + x)^2}
,
\end{equation}
\nop
where $x = r/r_s$, $\rho_s$ and $r_s = r_{vir}/c_{vir}$ are scaling
parameters of the density and radius, and $c_{vir}$ is the concentration parameter; 
and a non-isothermal $\beta$ model for the gas,

\begin{equation}  \label{E:RHO}
      \rho (r) =  { \rho_0  \over (1 + y^2)^{3 \beta /2} }
,
\end{equation}
\nop 
where $y = r/r_{core}$, and $\rho_0$ and $r_{core}$ are the central
density and scale radius for the intracluster gas. 
We derive the temperature of the gas from the equation of hydrostatic
equilibrium using numerical integration.
We assume an equation of state that of the ideal gas with $\gamma = 5/3$.
We adopt $r_{core} = 0.12$ \RVIR, and $\beta=1$, which are consistent
with our analysis of relaxed clusters of galaxies drawn from 
cosmological numerical simulations (see \citealt{Molnet10ApJ723p1272}). 
Assuming a value for $M_{tot}$ and the gas mass fraction of 0.14,
we derive $R_{vir}$, $\rho_s$, $r_s$, and $\rho_0$. 
We handle the small fraction of baryonic matter in galaxies along with the DM
since they both may be assumed to be collisionless for our purposes.
With this assumption, our DM particles 
also represent baryonic matter locked up in galaxies.  
The number of DM particles at each cell in the simulations is
determined by the density and the total number of particle (we used 5 million particles).
The amplitude of the velocity of each individual DM particle is derived by 
sampling a Maxwellian distribution (assuming a local Maxwellian approximation)
with dispersion, as a function of radius, derived from the Jeans equation 
assuming an isotropic velocity dispersion \citep{LokasMamon2001MNRAS321}.
We derive the directions of the velocity vector of the DM particles 
assuming an isotropic distribution.
For a detailed description of the setup of our FLASH simulations of merging galaxy clusters,
see \cite{Molnaret2012ApJ748}.

%
%
\begin{figure}[t]
\centerline{
\includegraphics[width=.50\textwidth]{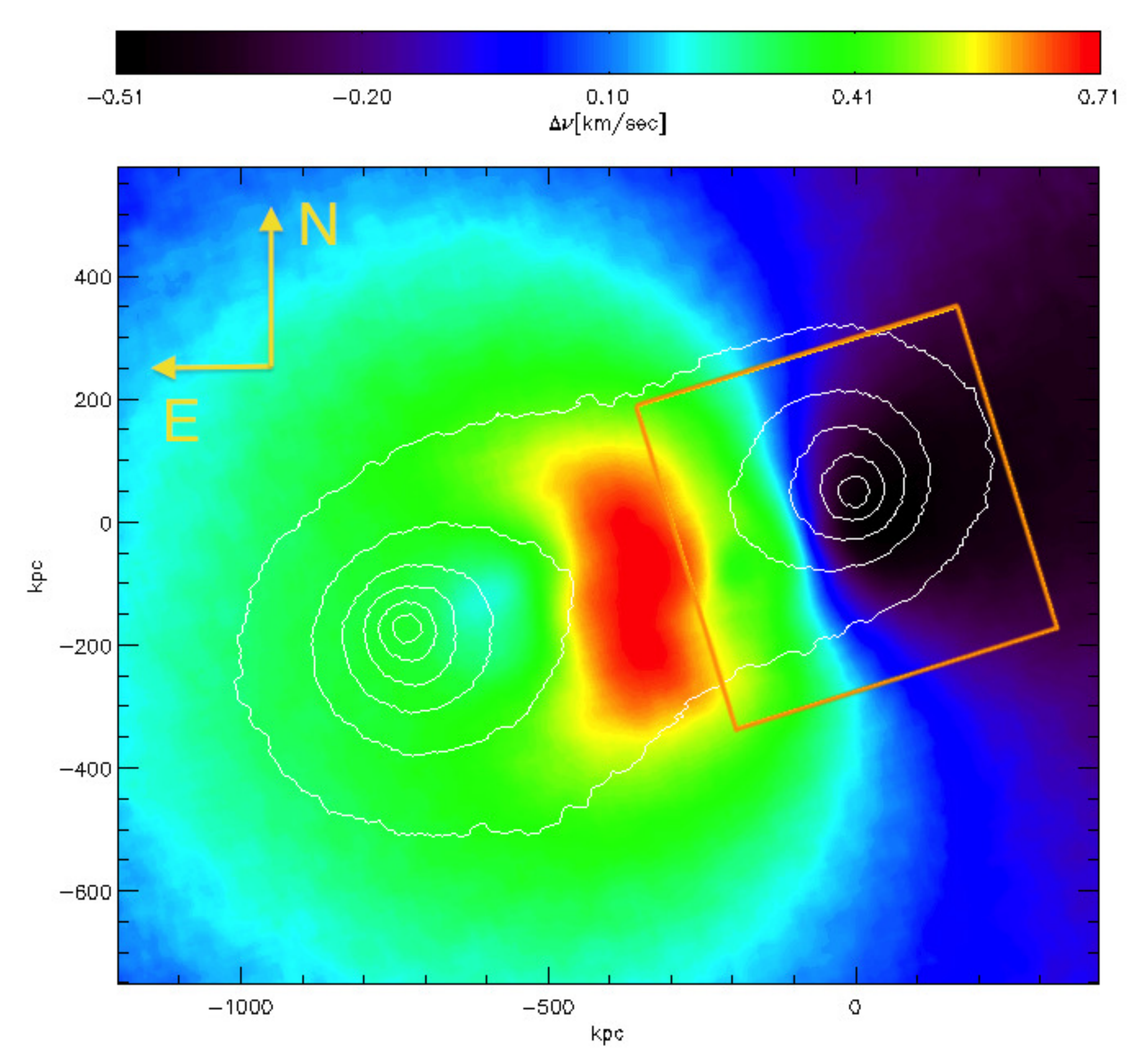}
}
\caption{
\protect
 Same as Figure~\ref{F:SIM_KAPVEL}, but for the frequency shift due to 
 gravitational lensing and tangential motion of the main component and the bullet 
 using Equation~\ref{E:DKMS12} with deflection angles and velocities 
 derived from our best fit simulation.
 The color bar on top shows the scale of the frequency shift in velocity units (\KMSEC).
 The white contours depict the convergence (shown in Figure~\ref{F:SIM_KAPVEL}) using                
 levels $\kappa$ = 0.2 to 1.2 with spacings of 0.2.                   
\label{F:SIM_DVELKMS}
}
\end{figure} 

Our main goal is to determine the velocity field of the dark matter component 
at the central part of the the bullet using \FLASH\ simulations which reproduce 
the observed main features of the mass distribution in the bullet cluster.
We carry out a set of simulations with different initial masses and concentration 
parameters but with infall velocity and impact parameter fixed at 3000 \KMSEC\ 
and 150 kpc, as suggested by the best model of \cite{Mastropietro08}.  
From our simulations we choose the snapshot at the same phase of the collision
as observed, i.e., when the two DM centers are 800 kpc apart.
From this snapshot we derive the projected mass from the profiles for both                
components, and compare both the mass at the center with the result from our              
strong lensing analysis, and compare the mass profiles with results from the              
lensing analysis of \cite{Bradac06}, as shown in Figure~\ref{F:MASSCUMUL}.               
We find that initial virial masses of M$_1$ = 9.0\TMSUNFOUR,
M$_2$ = 5.5\TMSUNFOUR, and concentration parameters $c_{vir1} = 6$, and
$c_{vir2} = 6$ provide a good agreement with the lensing results, 
although our mass profile of the bullet is shallower at the center than that from
our strong lensing analysis (see Figure~\ref{F:KAPPACONT}), and the mass profiles of 
both components are shallower than those based on lensing observations of \cite{Bradac06},
as we can see from Figure~\ref{F:MASSCUMUL}. 
Larger mass for the main component, which could provide a better fit for larger radii 
would yield too high a density at the center and would thus overestimate the central mass, 
which we avoid since we are mainly interested in the central cluster regions.
Since we underestimate the masses of the two components, we also 
underestimate the velocity increment of the bullet due to the infall 
at the phase of the observation; therefore our results for the frequency changes 
due to the moving cluster effect should be considered as lower limits.

The best fit model of the most successful numerical simulations of the bullet cluster, 
the SPH simulations of \cite{Mastropietro08}, assumed a mass ratio of 6.25 which is 
much larger than ours. However, their model produces a mass profile for the main 
component close to ours (slightly higher), but their result for the bullet component is 
only a half of the observed (Figure 6 of Mastropietro \& Burkert). 
Mastropietro \& Burkert also note that, based on the mass profiles derived from lensing, 
the mass of the bullet seems to be a significant fraction of the total mass, 
and the mass ratio should be lower than even the lowest mass ratio they considered, 3:1
(and much lower than 10:1 suggested by \citealt{Barrena2002AA386};
however, their result might not be very reliable because they used galaxy redshifts assuming 
relaxed halos and had only seven galaxies for the bullet component).
Mastropietro \& Burkert rejected models with lower mass ratios because in 
their SPH simulations the gas of the main component is disrupted to the degree that the 
X--ray peak of the main cluster disappears. In our AMR simulations with lower mass ratios 
we find that the main cluster gas is not disrupted (Molnar et al. 2013, in preparation).

%
%
 \begin{table}[t]
 \centering
 \caption{Frequency shifts for multiple systems due to the moving lens effect}
 \begin{tabular}{@{ }lccc@{ }}
 \toprule 
 \multirow{3}*{Name} &   \multicolumn{3}{c}{Frequency shift in \KMSEC}               \\
                                  &   \multicolumn{3}{c}{Rotation angle}                                   \\ 
                                  &  $\varphi$ = 0\DEG & $\varphi$ = +30\DEG  &  $\varphi$ = -30\DEG  \\
\midrule   
    G1      &   -0.172  &    0.005   &   -0.177     \\ 
    G2      &  -0.174   &    0.146   &   -0.253    \\ 
    G3      &   -0.130  &    0.290   &   -0.279     \\ \hline
    H1      &   -0.214  &    0.017   &   -0.226     \\ 
    H2      &   -0.198  &    0.174   &   -0.289       \\ 
    H3      &   -0.089  &    0.468   &   -0.316      \\ \hline
    I1        &   -0.122  &    0.395  &   -0.316        \\ 
    I2        &   -0.194  &    0.210  &   -0.300      \\ 
    I3        &   -0.161  &    0.300  &   -0.314      \\ \hline
    J1       &    0.016  &    0.386  &   -0.178      \\ 
    J2       &    0.016  &    0.272  &   -0.124       \\ 
 \bottomrule
 \end{tabular}
 \label{T:TABLE1} 
\tablecomments{
 Expected frequency shifts for multiple images of background galaxies 
 generated by the moving cluster effect near the
 bullet in the bullet cluster expressed in velocity units. 
 The rotation angle, $\varphi$, is measured from the base line defined 
 by the alignment of the simulations with the bullet cluster image. \\
 \hspace*{8pt}$^a$ Name identification; the same as in \cite{Bradac06}
 }
\medskip
\end{table}

Using our simulations of the best fit mass distribution model
at the phase of the bullet cluster (distance between the two
DM centers equal to 120 kpc), we generate 
DM surface mass density maps $\Sigma(\theta)$, 
and velocity field maps for the two mass components.
We integrate the total (DM and gas) density along the LOS
assuming that the collision is in the plane of the sky
(good approximation for the bullet cluster, e.g., \citealt{Mastropietro08}) 
to obtain the mass surface density images and convert the density map 
to a convergence map using the physical parameters relevant to the
bullet cluster. We determine the average velocity at each position
using the median of all bulk-velocity motions along the LOS.
We show the color image of the convergence for this phase of the collision in 
Figure~\ref{F:SIM_KAPVEL}.
The alignment of the image in RA and DEC is the same as that of the bullet cluster.
The image coordinates are given in physical units (kpc) centered on the peak in the 
surface mass density corresponding to the bullet.  
In this simulated image, the bullet, moving from left to right, just passed the core of the 
main cluster (first core passage) from the North.
 The white arrows represent the bulk motion of the mass within the cluster. 
 The length of the arrows show the amplitude of the velocities
 (with a maximum of 4000 \KMSEC).
Dominated by the bulk flow of the DM component of the bullet,
we find that the velocity field is smooth in the vicinity of the
bullet DM center (Figure~\ref{F:KAPVEL}), as we expected.

%
%

%
%
\section{Results and Discussion}
\label{S:RESULTS}

Here we first derive the expected frequency shifts due to the moving cluster
effect using the velocity and deflection fields determined from our 
simulations of the bullet cluster.
In our case, Equation~\ref{E:WSFRSHIFT} 
reduces to the sum of the two components, and the frequency shift field
from simulations, expressed in velocity units, can be calculated as
\begin{equation}  \label{E:DKMS12}
     V^{sim} (x,y) = \bv_{T1}  \cdot \balpha_1  +  \bv_{T2}  \cdot \balpha_2 
,
\end{equation} 
where the 2D vector fields $\bv_{T1,2}$ and $\balpha_{1,2}$
represent the tangential velocities and the deflection angle of the 
main cluster (1) and the bullet (2).
The frequency shift field expected over the surface of the cluster are shown in
Figure~\ref{F:SIM_DVELKMS}. A dipole pattern for the fast--moving
component, the bullet, can be clearly seen on the right hand side of
the image. The difference between the maximum and minimum 
frequency shift due to the tangential motion of the bullet is 1.2 \KMSEC.

%
%
 \begin{table}[t]
 \centering
 \caption{Large frequency shift differences due to the moving cluster effect}
 \begin{tabular}{@{ }lcccc@{ }}
 \toprule 
 \multirow{3}*{Name} &                     \multicolumn{4}{c}{Frequency shift difference in \KMSEC}               \\
                             &  \multicolumn{3}{c}{Moving lens effect} &    Moving observer \\ 
                 &  $\varphi$ = 0\DEG & $\varphi$ = +30\DEG  &  $\varphi$ = -30\DEG   & no $\varphi$ dependence \\
\midrule                                          
  G3-G1  &     -       &  0.284  &  0.103 &      -0.020    \\ 
  G3-G2  &     -       &      -      &     -       &       -           \\ 
  G2-G1 &     -       &      -      &     -       &     -0.009     \\ \hline
  H3-H1  &  0.125  &  0.451  &    -       &      -0.035     \\ 
  H3-H2  &  0.109  &      -      &    -        &     -0.024    \\ \hline
  I2-I1         &     -        &  0.185  &    -         \\ \hline
  J2-J1        &      -       &  0.115  &     -        \\ 
 \bottomrule            
 \end{tabular}
 \label{T:TABLE2} 
 \tablecomments{
  Same as Table~\ref{T:TABLE1}, but for the differences in 
  frequency shifts greater than 0.100 \KMSEC\
  for images of the same multiply-lensed background galaxy.
 }
\medskip
\end{table}

Our simulations do not include structures smaller than the 
two primary mass components, approximating the bullet cluster system.
We also ignore the often significant contribution of member galaxies to the lensing
deflection field in our idealized simulations. So here we make the
frequency shift calculation more realistic by deriving the lensing
deflection field using the actual observed multiple images with our
lens model for which the observed member galaxies are incorporated
(see Section \ref{S:LENSING}).
Details of the velocity field are not that important
as this field is inherently smooth, being dominated by the bulk motions of the DM 
components of the two main clusters, and therefore this field
is reasonably well mapped, as shown above.
In the vicinity of the bullet the observed deflection field is mainly due to the DM 
component of the bullet, and both the velocity of the DM and the deflection field 
generated by the main cluster are negligible.
In this case, Equation~\ref{E:WSFRSHIFT} reduces to Equation~\ref{E:DELNU},
and we can approximate the expected frequency shift due to
the moving cluster effect as
\begin{equation}  \label{E:DKMSAPPROX}
     V(x,y) = \bv_{T} (x,y) \cdot \balpha (x,y)
,
\end{equation} 
where the velocity field of the bullet, $\bv_{T}(x,y)$ is taken from our FLASH 
simulations (Section~\ref{S:FLASH}) and the deflection field, $\balpha (x,y)$, 
is from our strong lensing analysis (Section~\ref{S:LENSING}).
In Figure~\ref{F:KAPVEL} we show the velocity field from our simulations
(white arrows) oriented to match the alignment of the ACS/HST images of the bullet
cluster plotted over the convergence field we derive
from strong lensing.  The map of the frequency shifts is plotted in
units of \KMSEC\ in Figure~\ref{F:DVELKMS} (with 
white contours overlaid representing the convergence field 
shown in Figure~\ref{F:KAPVEL} in colors).

From the frequency shift field, $V(x,y)$, shown in Figure~\ref{F:DVELKMS}
we now evaluate the expected frequency shifts at the positions of the 
multiply-imaged background galaxies, $V_i = V(\btheta_I^i$), as
identified by \cite{Bradac06}. 
We show the list of the frequency shifts due to the moving cluster effect 
in velocity units in Table~\ref{T:TABLE1}.   
The first column is the ID designation of the
system as in Bradac et al.  The second column represents the frequency
shifts we obtain when we align the velocity field derived from
simulations with the deflection angle from strong lensing
observations.  The remaining 2 columns illustrate possible systematic errors in the 
frequency shifts allowing $\pm 30$\DEG\ rotation angles, $\varphi$, 
relative to the rotation angle chosen for the alignment of simulations and observations.

In principle, we can measure the relative frequency shift between any
of the counter images belonging to each multiply-lensed background
galaxy.  These relative frequency shifts are shown in
Table~\ref{T:TABLE2} for systems that we expect to exceed 0.100
\KMSEC.  From this table we find that two triply--imaged systems (G
and H) having the largest relative frequency shifts about 0.5 \KMSEC\,
depending on the orientation of the DM velocity field. 
However, we consider these as underestimated relative shifts, 
since we find that at this infall velocity (3000 \KMSEC),
our preliminary AMR simulations, aimed to explain the mass 
surface distribution, and the X--ray and SZ morphology, have
insufficient Ram pressure to reproduce the 
observed displacement between the DM center of
the lower mass bullet component and the bullet gas.
Instead we find that an initial relative velocity of 4500 \KMSEC\ is required
between the cluster components in order to create a distinct bullet with a 
sizable offset between the gas and dark mater, as observed.
This larger impact velocity would result in frequency
shifts of about 50\% larger than those based on an impact velocity of
3000 \KMSEC\ we are using in our analysis.  As a result, our estimates
for the frequency shifts due to tangential motion of the bullet
cluster can be considered as conservative lower estimates.

As a consistency check we estimate the relative frequency shifts due to the
tangential motion of the observer, $\bv_T^{\,O}$.
In the case of the bullet cluster, the relative frequency shift
for the $i$th lensed image, at the image position, $\btheta_I^i$, 
expressed in velocity units using the second term in Equation~\ref{E:WSFRSHIFT},
becomes
\begin{equation}    \label{E:FR_OBS}
   V_{{\rm X}i} = - \frac{ D_{LS}(z, z_{\rm X}) } { D_{OS}(z_{\rm X}) } \,
                                                \bv_T^{\,O} \cdot \balpha (\btheta_I^i)
,
\end{equation} 
where X is the index for the lensed background galaxy, X = G, H, I, or J, and
$z_{\rm X}$ is the corresponding redshift, $z_{\rm G}$ = 1.3,
$z_{\rm H}$ = 1.9, $z_{\rm I}$ = 2.1, $z_{\rm J}$  = 1.7 \citep{Bradac06}.
We obtain $\bv_T^{\,O}$ by projecting our Heliocentric velocity relative to the CMB,
$\bv^{\,O}$, to the plane of the sky of the bullet cluster.
Using RA = 6h 58m 37.9s; DEC = -55\DEG\ 57\arcmin\ 0\arcsec\ for the position
of the bullet cluster and 
$v^{\,O}$ = 369.0 \KMSEC\ pointing toward 
RA = 11h 11m 43s; DEC = -6\DEG\ 55\arcmin\ 37\arcsec\ for the velocity of the
observer \citep{HinshawET2009ApJS180}, 
we obtain $v_T$ =  345.672 \KMSEC, with a rotation angle of 
103.\DEG9 relative to the velocity of the bullet in the plane of the sky. 
We find that the largest relative frequency shift due to the observer's motion 
is 35 \MSEC\ (for the H3-H1 pair), and in most cases, these frequency shifts are 
around 20 \MSEC, thus we conclude that these can be safely neglected 
(see Table~\ref{T:TABLE2}).

As we have shown above such predicted frequency shifts are within the
range of the ALMA array for molecular CO(1--0) emission and near the
resolution limit in the optical-IR using X-shooter on the VLT. Ideally
the velocity resolution should be sufficient to allow for the definition
of a detailed velocity profile over the emission lines detected for
each component of a given multiply-lensed galaxy, so that repeatable
velocity substructure can be identified and used to help establish the
small relative velocity shifts induced by the tangential motion of the lens.
There is a great advantage in using systems with more than 2 multiple images in 
that the direction of the motion of the cluster mass can be solved for  
(see our Equation~\ref{E:SHIFTS3}) without the need for models of the DM motion.

%
%
\section{Final Comments}
\label{S:FINAL}

  High speed motions recently inferred from large scale shocked gas 
  and offsets between the positions of DM and gas in
  colliding clusters seem surprising in the context of standard
  cosmological models. The bullet cluster, which provided the first
  example for this phenomenon, is claimed to be in grave conflict with 
  the concordance \LCDM\ model, where no such high speed encounters 
  of such massive clusters is predicted in even the largest simulations. 
  However, as we discussed in the introduction, several other merging
  clusters exist with inferred large relative impact velocities.

The main motivation for considering the challenging observations
  proposed here lies in the ability to measure the motion of dark
  matter directly from observed frequency shifts due to the moving cluster effect.
 The moving cluster effect investigated here is generated by the motion of the 
 gravitational potential and therefore its importance lies in the direct relation to
 the motion of the dominant DM, unlike the kSZ effect,
 for which gas modeling is required.
  The bulk motion of the DM is smooth even in merging clusters,
  as we demonstrated using our simulations, therefore it is 
  much easier to model this motion unlike that of the gas which has a 
  complicated velocity field due to non-gravitational interactions.

We have examined the moving cluster effect on the 
relative frequency shifts
between multiply-imaged 
background galaxies in the bullet cluster
from detailed modeling of the velocity and gravitational potential fields.
Our results are shown in Table~\ref{T:TABLE2}. 
We have found two triply-imaged systems having the largest relative frequency
shift of about 0.5 \KMSEC\ depending on the orientation of the dark-matter velocity field. 
We have found that at this infall velocity, 
our AMR simulation has insufficient Ram pressure to reproduce the observed 
displacement between the dark-matter center of the bullet component 
and the bullet gas, which we find requires initial relative velocities of as large as 
4500 \KMSEC\ for which a distinct bullet is created with a relatively large offset 
between the gas and DM, as observed. 
This larger impact velocity would result in
frequency shifts of about 50\% larger than those 
based on the impact velocity of 3000 \KMSEC\ we assumed in our analysis. 
As a result, our estimates for the frequency shifts due to tangential motion of the 
bullet cluster are conservative.
These predicted frequency shifts are 
within the range of the ALMA array for molecular CO(1--0) emission 
and near the resolution limit in the optical-IR using X-shooter on the VLT.

  This proposed method can be extended to smaller frequency shifts
  using many independent lines of the Lyman-$\alpha$ forest towards distant QSOs
  lensed by more typical 
  clusters of galaxies, complementary to line-of-sight peculiar motions derived from the
  kinematic SZ effect, so that the 3D peculiar motions of
  individual clusters may be reconstructed from direct measurements.

%
%
\acknowledgments

We thank the referee for valuable comments and suggestions which improved the 
presentation of our results.
The code FLASH used in this work was in part developed by the
DOE-supported ASC/Alliance Center for Astrophysical Thermonuclear
Flashes at the University of Chicago.  We thank the Theoretical
Institute for Advanced Research in Astrophysics (TIARA), Academia Sinica, 
for allowing us to use its high performance computer facility for our
simulations.  
K. U.  acknowledges partial support from the National Science Council of Taiwan 
(grant NSC100-2112-M-001-008-MY3) and from
the Academia Sinica Career Development Award.
A. Z. is supported by contract research ``Internationale Spitzenforschung II/2-6'' of 
the Baden W\"urttemberg Stiftung.

%
%
\bibliographystyle{apj}


\end{document}